\documentclass[aps,twocolumn,10pt,longbibliography,superscriptaddress]{revtex4-1}

\usepackage{amsmath}
\usepackage{amsfonts}
\usepackage{latexsym}
\usepackage{graphicx}
\usepackage{color}
\usepackage{bm}
\usepackage{float}
\usepackage{hyperref} 
\usepackage[all]{hypcap}

\newcommand{\bra}[1]{\langle #1|}
\newcommand{\ket}[1]{|#1\rangle}

\begin{document}
\title{Characterizing an Entangled-Photon Source with Classical Detectors and Measurements}

\author{Lee A. Rozema}
\thanks{Corresponding author: lee.rozema@univie.ac.at}
\affiliation{Centre for Quantum Information \& Quantum Control and Institute for Optical Sciences, Department of Physics, 60 St. George St., University of Toronto, Toronto, Ontario, Canada M5S 1A7}
\affiliation{Faculty of Physics, University of Vienna, Boltzmanngasse 5, A-1090 Vienna, Austria}

\author{Chao Wang}
\affiliation{Centre for Quantum Information \& Quantum Control and Institute for Optical Sciences, Department of Physics, 60 St. George St., University of Toronto, Toronto, Ontario, Canada M5S 1A7}

\author{Dylan H. Mahler}
\affiliation{Centre for Quantum Information \& Quantum Control and Institute for Optical Sciences, Department of Physics, 60 St. George St., University of Toronto, Toronto, Ontario, Canada M5S 1A7}

\author{Alex Hayat}
\affiliation{Centre for Quantum Information \& Quantum Control and Institute for Optical Sciences, Department of Physics, 60 St. George St., University of Toronto, Toronto, Ontario, Canada M5S 1A7}
\affiliation{Canadian Institute for Advanced Research, Toronto, Ontario M5G 1Z8, Canada}
\affiliation{Department of Electrical Engineering, Technion, Haifa 32000, Israel}

\author{Aephraim M. Steinberg}
\affiliation{Centre for Quantum Information \& Quantum Control and Institute for Optical Sciences, Department of Physics, 60 St. George St., University of Toronto, Toronto, Ontario, Canada M5S 1A7}
\affiliation{Canadian Institute for Advanced Research, Toronto, Ontario M5G 1Z8, Canada}

\author{John E. Sipe}
\affiliation{Centre for Quantum Information \& Quantum Control and Institute for Optical Sciences, Department of Physics, 60 St. George St., University of Toronto, Toronto, Ontario, Canada M5S 1A7}

\author{Marco Liscidini}
\affiliation{Dipartimento di Fisica, Universit{\'a} degli Studi di Pavia, via Bassi 6, I-27100 Pavia, Italy}

\begin{abstract}
Quantum state tomography (QST) is a universal tool for the design and optimization of entangled-photon sources.
It typically requires single-photon detectors and coincidence measurements.
Recently, it was suggested that the information provided by the QST of photon pairs generated by spontaneous parametric down-conversion could be obtained by exploiting the stimulated version of this process, namely difference frequency generation. In this protocol, so-called ``stimulated-emission tomography'' (SET), a seed field is injected along with the pump pulse, and the resulting stimulated emission is measured. Since the intensity of the stimulated field can be several orders of magnitude larger than the intensity of the corresponding spontaneous emission, measurements can be made with simple classical detectors. Here, we experimentally demonstrate SET and compare it with QST. We show that one can accurately reconstruct the polarization density matrix, and predict the purity and concurrence of the polarization state of photon pairs without performing any single-photon measurements.
\end{abstract}

\maketitle

Quantum information is an important emerging technology \cite{nielsen_quantum_2000}, and entanglement is its essential ingredient. It
plays a vital role in tasks such as quantum computation \cite{Raussendorf_measurement_2003, walther_experimental_2005}, quantum
metrology \cite{boto_quantum_2000,mitchell_super-resolving_2004}, and
quantum key distribution \cite%
{Ekert_quantum_1991,ursin_entanglement_2007,Jennewein_quantum_2000}, and so
progress in the development of high-quality entanglement sources is of
central importance for advances in quantum information technology.

For any source of entangled states to be useful, it must be characterized.
The standard method for doing this is quantum state tomography (QST) \cite%
{james_measurement_2001}. In principle, QST provides a complete description of
the quantum state, from which one can evaluate the suitability of a source
for any proposed application. But it is well-known that QST is a
resource-intensive task. The quality of the tomographic estimate depends on
the amount of data that one is able to acquire \cite%
{gill_state_2000,Okamoto_experimental_2012,mahler_adaptive_2013} and analyse 
\cite{gross_quantum_2010}, with more data typically resulting in a
higher-quality estimate.

In this work, we investigate a particular physical system --
entangled-photon pairs generated via spontaneous parametric down-conversion
(SPDC) in a pair of BBO crystals -- and show how the corresponding
stimulated process, namely difference frequency generation (DFG), can be
used to reconstruct the polarization density matrix of the two-photon state
that arises in the spontaneous process. The signal generated by DFG can be
several orders of magnitude more intense than that observed in SPDC, making
it possible to estimate the quantum states produced by such sources very
rapidly and efficiently. \cite{Liscidini_Stimulated_2013}. This is
particularly important en route to the development of \textquotedblleft
on-chip\textquotedblright\ sources of entangled states \cite%
{horn_Monolithic_2012,orieux_direct_2013,horn_inherent_2013}; as in the case
of integrated electronic circuits, such sources are typically produced in
large numbers, and fast and efficient characterization procedures are
required. \ The approach of \textquotedblleft stimulated emission
tomography\textquotedblright\ (SET) we demonstrate here would also allow for
the full quantum characterization of sources with very low photon-pair
generation rates, for which characterization by conventional QST might not
even be feasible. \ While an experiment exploiting the relation between
stimulated and spontaneous emission to measure spectral correlations
between the spontaneously generated photons has been performed recently \cite%
{eckstein_direct_2013,Fang_fast_2014}, the full tomography of photon pairs using SET has
yet to be demonstrated. \ That is what we undertake here.

As test case for SET, we use our polarization-entangled photon source 
(an SPDC \textquotedblleft sandwich source\textquotedblright \cite{Kwiat_typeII_1995}) 
which is illustrated in Fig. 1a; here\
a pair of nonlinear birefringent crystals is mounted with their optic axes
orthogonal. \ In an idealized picture, when a diagonally polarized pump
pulse is incident the first crystal could produce pairs of horizontally
polarized photons in the signal and idler modes (taking $|V\rangle _{{p}%
}\rightarrow |HH\rangle _{s,i}$), or the second crystal could produce
vertically polarized photon pairs (taking $|H\rangle _{{p}}\rightarrow
|VV\rangle _{s,i}$). Thus, if the source \ -- i.e. the pump pulse, the
nonlinear crystals, and the collection optics -- is appropriately configured 
\cite{altepeter_Phase-compensated_2005,Rangarajan_optimizing_2009}, SPDC results in the generation of a
polarization-entangled state $(|HH\rangle _{s,i}+|VV\rangle _{s,i})/\sqrt{2}$.

\begin{figure}[tbp]
\centering
{\includegraphics[width=\linewidth]{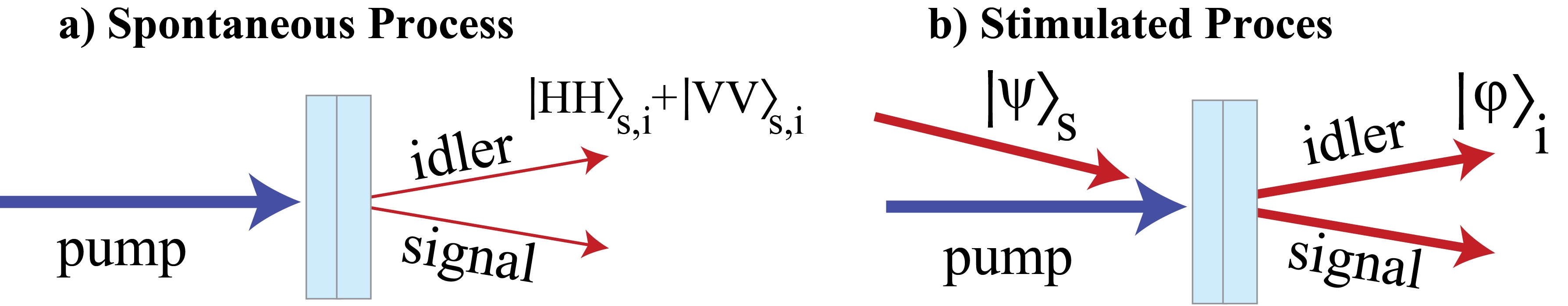}}
\caption{a) A simple cartoon
of a SPDC ``sandwich'' source which produces entangled-photon pairs. b) A
stimulated version of the SPDC sandwich source, now a seed beam is sent into
the signal mode to stimulate light into the idler mode.}
\end{figure}

To characterize this state using QST, the polarization of each photon is measured in several different bases.  
Experimentally, this is typically done by sending each photon to a set of waveplates and a polarizer (Fig. 2b), and estimating the probability that both photons are transmitted.
For example, if both polarizers are aligned in the horizontal direction we can estimate $\mathcal{P}_{H,H}$.
For two-photons, 16 such probabilities are sufficient to constrain the two-photon polarization state.
Thus, these measurements can be used to estimate the two-photon polarization density matrix \cite{james_measurement_2001};
see the Supplement for more details.

In SET, a strong seed beam is constructed to mimic the signal photons of the
pair that could be generated by spontaneous
emission (Fig. 1b), including all possible polarizations.
 Consequently, in the presence of the seed beam,
an idler beam is generated by DFG. Two of us predicted
earlier \cite{Liscidini_Stimulated_2013} that the biphoton wavefunction
characterising the pairs that would be emitted by SPDC acts as the response
function relating the idler beam to the seed beam in DFG. Thus, by changing
the polarization of a properly configured seed beam $\ket{\psi}_s$ and performing measurements on the
polarization of the stimulated idler $\ket{\phi}_i$, conclusions can be drawn about the
two-photon state that would be generated in the absence of the seed. For
example, by using a horizontally polarized seed beam and measuring the
stimulated beam in the horizontal basis, one can compute the probability of
detecting a horizontally polarized signal photon and a horizontally polarized
idler photon in a spontaneous experiment $\mathcal{P}_{H,H}$. In an ideal
situation, this probability is simply 
$\mathcal{P}_{H,H}={I_{H}^{\mathrm{stim}}}/{I_{H}^{\mathrm{seed}}}$,
where $I_{H}^{\mathrm{stim}}$ is the intensity of the horizontally polarized
simulated light and $I_{H}^{\mathrm{seed}}$ the intensity of the seed light.
This procedure can be repeated for other polarization combinations, yielding
the same information as QST.
Thus, SET obtains sufficient information to reconstruct the full 
two-photon polarization state.

The source and our implementation of QST and SET are shown in Fig. 2. Our
sandwich crystals are a pair of $1$ mm - thick BBO crystals. To generate
entangled photons, a \textquotedblleft temporal
compensation\textquotedblright\ crystal is placed before the sandwich
crystals to pre-delay one component of the pump polarization so that, by the
time the photon-pairs emerge from the sandwich crystals, the $\ket{H,H}_{s,i}$ pairs are temporally
indistinguishable from the $\ket{V,V}_{s,i}$ pairs. Our source is pumped with $\approx 200$ fs-long pulses 
which are centred at $400$ nm, with an average power of $500$ mW.
This pump light is generated by frequency-doubling $1.5$ W (average power) 
of $800$ nm light from a femtosecond Ti:Sapph laser with a 76 MHz repetition rate, using a $2$ mm-long BBO crystal. 
The $400$ nm pump pulse is focused into the crystal with a $15$ cm lens, resulting in a beam 
waist of $\approx 50$ $\mu$m in the crystal.

The collected signal and idler modes are defined by two single-mode fibres
(with a numerical aperture of $0.12$), a $10$ cm focal-length lens to
collimate the emission, and two $4$ mm aspherical lenses to focus the signal
and idler beams into the fiber. Light is collected from a spot size that is
about the same size as the focused pump, as prescribed by Ling et al. \cite%
{ling_absolute_2008}. \ Finally, each mode is filtered with a $10$ nm
spectral filter centred at $800$ nm. This produces approximately $15,000$
polarization-entangled photon pairs per second coupled in the signal and idler
modes, with a coupling efficiency (pairs/singles) of $\approx 15\%$. These
photons are then directed to a standard QST apparatus (see Fig. 2b), where
the photon pairs are detected with single-photon detectors,
and coincidence counts are registered with a homebuilt FPGA-based
coincidence circuit.  When the source is nominally configured to generate the
maximally-entangled state $(|HH\rangle _{s,i}+|VV\rangle _{s,i})/\sqrt{2}$,
QST yields the polarization density matrix shown in Fig. 3a, which has a fidelity
of $\approx 0.951$ with the maximally entangled state.

\begin{figure}[tbp]
\centering
{\includegraphics[width=\linewidth]{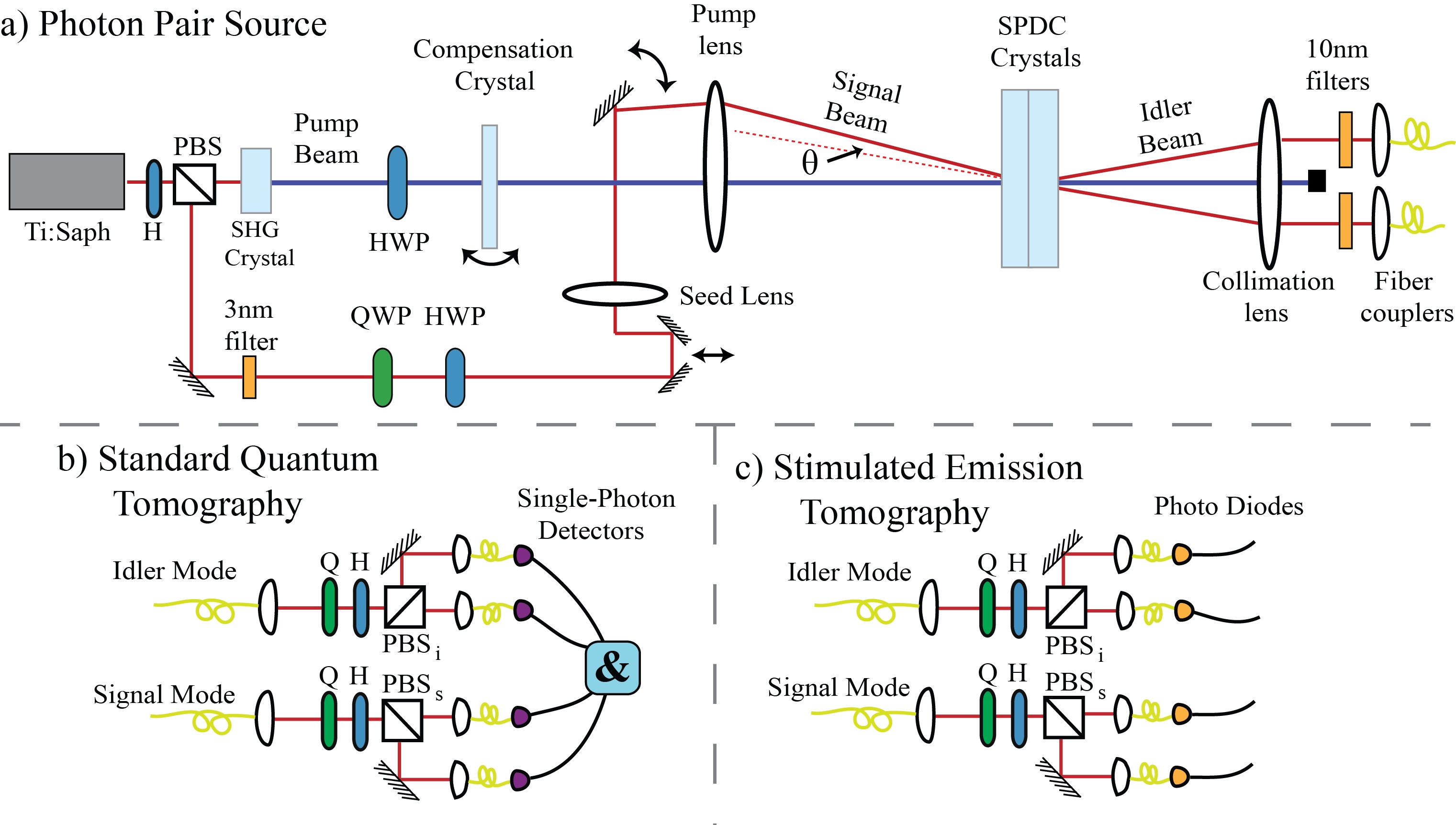}}
\caption{
a) A sketch of our entangled-photon source and the generation of our seed beam for stimulated
emission tomography. The signal and idler modes are coupled into a pair of
single-mode fibres, and sent to a tomography apparatus which can be
configured for either quantum or stimulated-emission tomography. b) The
standard quantum tomography apparatus. It consists of a pair polarization
measurements on each mode (implemented using waveplates and polarizing
beamsplitter cubes). For quantum state tomography, the output modes of each
cube are coupled to single-photon detectors and coincidences
between the various detector pairs are monitored. c) The stimulated emission
tomography apparatus. It is almost identical to quantum tomography, but now
photo diodes replace the single-photon detectors and there is no need for any coincidence
measurements. For both quantum and stimulated-emission
tomography, both ports are monitored to aid in normalizing the data to
correct for different detector efficiencies.}
\end{figure}

To perform SET, a seed field is constructed by diverting approximately $150
$ mW of the original $800$ nm pulsed Ti:Sapph light; in this configuration, the 
$400$ nm average pump power is also $150$ mW.
Note that this is not a strict application of SET as proposed
earlier \cite{Liscidini_Stimulated_2013}, since it makes use of a pulsed
earlier \cite{Liscidini_Stimulated_2013}, since it makes use of a pulsed
seed and not a CW seed. This can introduce errors in the determination of
the polarization density matrix when the biphoton wavefunction depends
strongly on the photon energy within the seed pulse bandwidth. To compensate
for this, we used a $3$ nm spectral filter to lengthen the seed pulse.
Finally, since the pump and the seed pulse have different frequencies, they
experience different spatial walk-offs. To minimize this effect the seed beam
waist is made much larger than the pump beam waist (approximately $1000$
$\mu$m), so that good spatial overlap is maintained as both beams traverse
the nonlinear crystals. Note that if the \textquotedblleft seed
lens\textquotedblright\ (Fig. 2a) is removed, then the seed pulse would have
a waist of $50$ $\mu$m in the crystal; in this configuration the different
spatial walk-offs become relevant. As we discuss in Section 3 of the
Supplement, this can lead to polarization dependent losses, which can
introduce errors in the SET reconstruction. When the $1000$ $\mu$m seed
beam is temporally and spatially overlapped with the pump pulse in the SPDC
crystals, we observe a stimulated idler output power of approximately $100$ $%
\mu$W coupled into the single-mode fibre. The polarization measurements for
SET are then performed in the same apparatus used for QST, but with the
single-photon detectors replaced by photo-diodes (Fig. 2c).

\begin{figure}[tbp]
\centering
{\includegraphics[width=\linewidth]{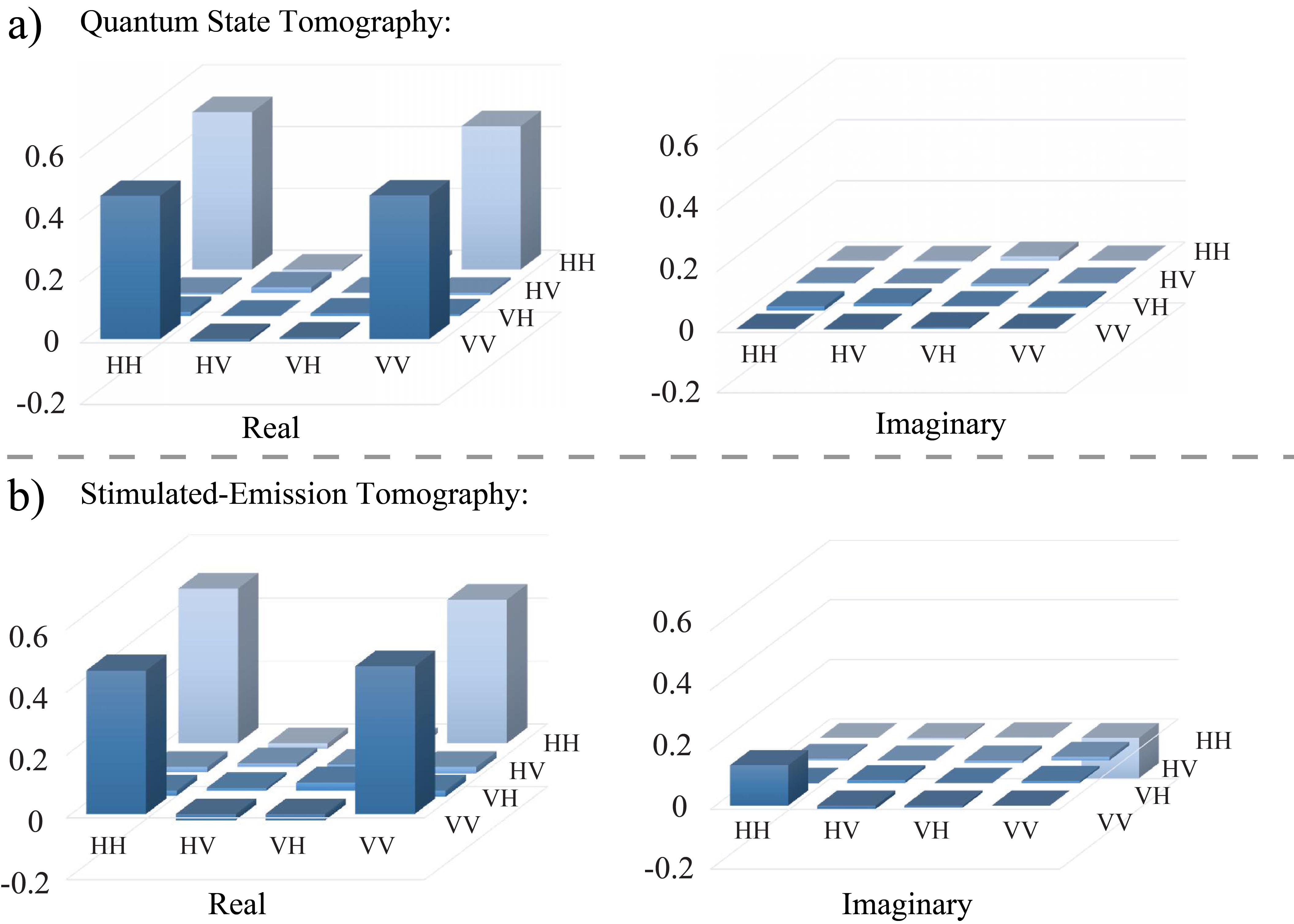}}
\caption{
Two-photon density
matrices reconstructed by (a) standard quantum state tomography and (b)
stimulated emission tomography (SET). The SET
reconstruction is based on the single-photon density matrices shown in Table 1 of the Supplement.
The density matrices reconstructed using the two methods have a fidelity of
0.963 with each other. }
\end{figure}

We prepare the seed signal pulse in six different polarization states, and,
for each seed configuration, we measure the intensity of the stimulated
idler in the same six different polarization states (see
Table 1 of the Supplement). These data are fed into
our modified least-squares fitting algorithm (see Section 1 of the
Supplement) to reconstruct the quantum state. When the source is nominally
configured to generate the maximally entangled state $(|HH\rangle
_{s,i}+|VV\rangle _{s,i})/\sqrt{2}$, SET yields the polarization density
matrix shown in Fig. 3b, which has a fidelity of $\approx 0.939$ with the 
maximally-entangled state.

The results of QST and SET are close but significantly different; the
quantum fidelity of one with respect to the other is $0.963$ \cite%
{jozsa_fidelity_1994}. The difference arises because the phase between $|HH\rangle _{s,i}$ and $|VV\rangle _{s,i}$ extracted by SET and QST disagree
by $0.289$ rad; this is manifest in the larger imaginary components in the
SET prediction. Were the phases artificially set to be identical, the fidelity between
the two estimates would increase to $0.982$.

\begin{figure}[tbp]
\centering
{\includegraphics[width=\linewidth]{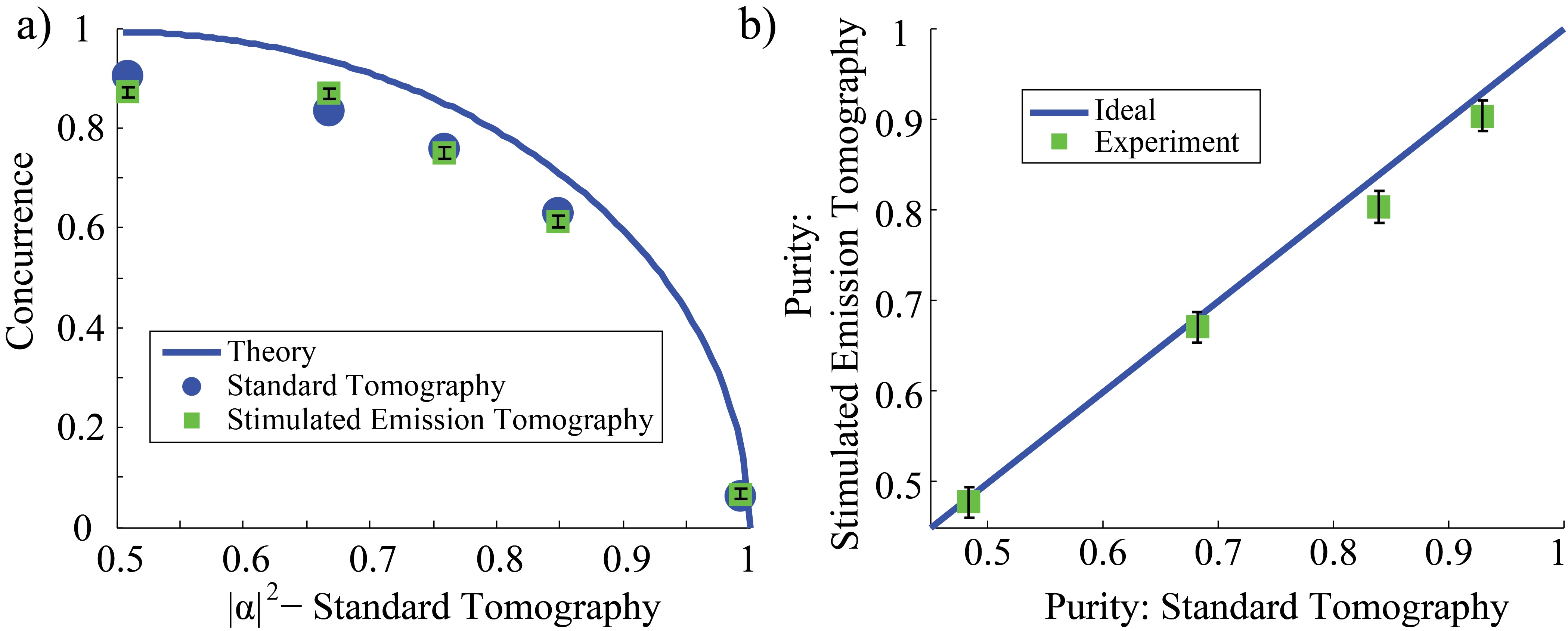}}
\caption{a) A plot of the concurrence
versus $|\protect\alpha|^2$; $|\protect\alpha|^2$ was extracted from
 quantum state tomography (QST). These data were taken for entangled states of
the nominal form: $\protect\alpha|HH\rangle+\protect\beta|VV\rangle$. The green
squares were extracted from stimulated-emission tomography (SET), the blue circles from
are the QST, and the blue curve is
a simple theory calculation assuming perfectly pure states. b) A plot of the purity extracted from SET
versus the purity extracted from standard QST. For these
data $|\protect\alpha|^2\approx 0.5$, and the compensation crystal in the
source was systematically misaligned to reduce the purity.}
\end{figure}

We will return to the disagreement between QST and SET below, but first note that measures
of entanglement (such as the concurrence) are not sensitive to the phase
between $|HH\rangle _{s,i}$ and $|VV\rangle _{s,i}$. Hence, we should expect that QST
and SET would predict essentially the same amount of entanglement. \ 

To confirm this, in one set of experiments we prepare states of the form $\alpha
|HH\rangle _{s,i}+\beta |VV\rangle _{s,i}$ and vary $|\alpha |$ from $1/%
\sqrt{2}$ to 0 (by rotating the pump polarization with a half waveplate).
This gradually decreases the concurrence of the source (while keeping the state
approximately pure). We perform both SET and QST on states in this range;
the concurrences predicted by both techniques are plotted versus $|\alpha|^{2}$ in Fig. 4a. 
Since the concurrence is independent of the phase between $|HH\rangle _{s,i}$ and $|VV\rangle_{s,i}$, the QST and SET results agree extremely well.

The most common problem plaguing polarization-entanglement sources is a reduced coherence
between $|HH\rangle _{s,i}$ and $|VV\rangle _{s,i}$, which results in a loss
of entanglement \cite{Rangarajan_optimizing_2009}. 
{Therefore, we performed a second set of experiments, showing that SET can characterize this loss.}
In sandwich sources pumped with an ultra-fast laser, such
as ours, the entanglement is often reduced by imperfect temporal compensation; the $%
|HH\rangle _{s,i}$ and $|VV\rangle _{s,i}$ photons are emitted in
distinguishable temporal modes, so that ignoring these temporal modes
effectively decoheres the state. To see how QST and SET capture this loss of
polarization entanglement, we initially set (using QST) our source to
produce the nominal state $(|HH\rangle _{s,i}+|VV\rangle _{s,i})/\sqrt{2}$
with high purity, and then began to misalign the compensation crystal by
rotating it in $10^{\circ }$ increments, finally removing it
altogether, to reduce the purity of the states. Both QST and SET were
performed on these states, and the purities yielded by the two techniques are in
good agreement, as shown in Fig. 4b. \ Again, this agreement is independent
of the actual phase between $|HH\rangle _{s,i}$ and $|VV\rangle _{s,i}$.

We now return to the disagreement in the phase between QST and SET. \ A
deeper investigation allowed us to establish that the phase between $%
|HH\rangle _{s,i}$ and $|VV\rangle _{s,i}$ extracted by SET is a function of
the incidence angle of the seed beam ($\theta $ in Fig. 2) as illustrated in
Fig. 5. In particular, it varies by $\approx 0.312$ rad per mrad
deviation of the seed beam (extracted from the fit shown as a dashed line).
This is simply because the spontaneously generated photons can be emitted
at different angles, and the nonlinear crystals are birefringent.
The magnitude of our experimentally-measured angle-dependent phase agrees well with the
theory of Altepeter et al. \cite{altepeter_Phase-compensated_2005}, which
predicts $0.461$ rad per mrad for our crystals.

In our QST experiment, pairs are collected over a transverse momentum range $%
\Delta k_{f}=\lambda /(\pi \omega _{f})\approx $ $5$ mrad, which can be
estimated by observing that the single-mode fibers collect from a spot size
of $\omega _{f}\approx 50$ $\mu$m in the crystal; the angular
phase-matching bandwidth of our SPDC crystals is $\approx 3.5$ mrad \cite%
{SNLO}. \ Thus, the polarization matrix determined by QST is itself the
results of averaging over the emission angles, and the identification of a
state (as in Fig. 3a) such as $(|HH\rangle _{s,i}+|VV\rangle _{s,i})/\sqrt{2}
$ indeed is truly only nominal; in fact, there is entanglement between
polarization degrees of freedom and emission angle. \ This is a
well-known result in bulk down-conversion experiments where one often collects pairs over broad
range to increase the detection rate \cite{altepeter_Phase-compensated_2005,Rangarajan_optimizing_2009}.

\begin{figure}[t]
\centering
{\includegraphics[width=\linewidth]{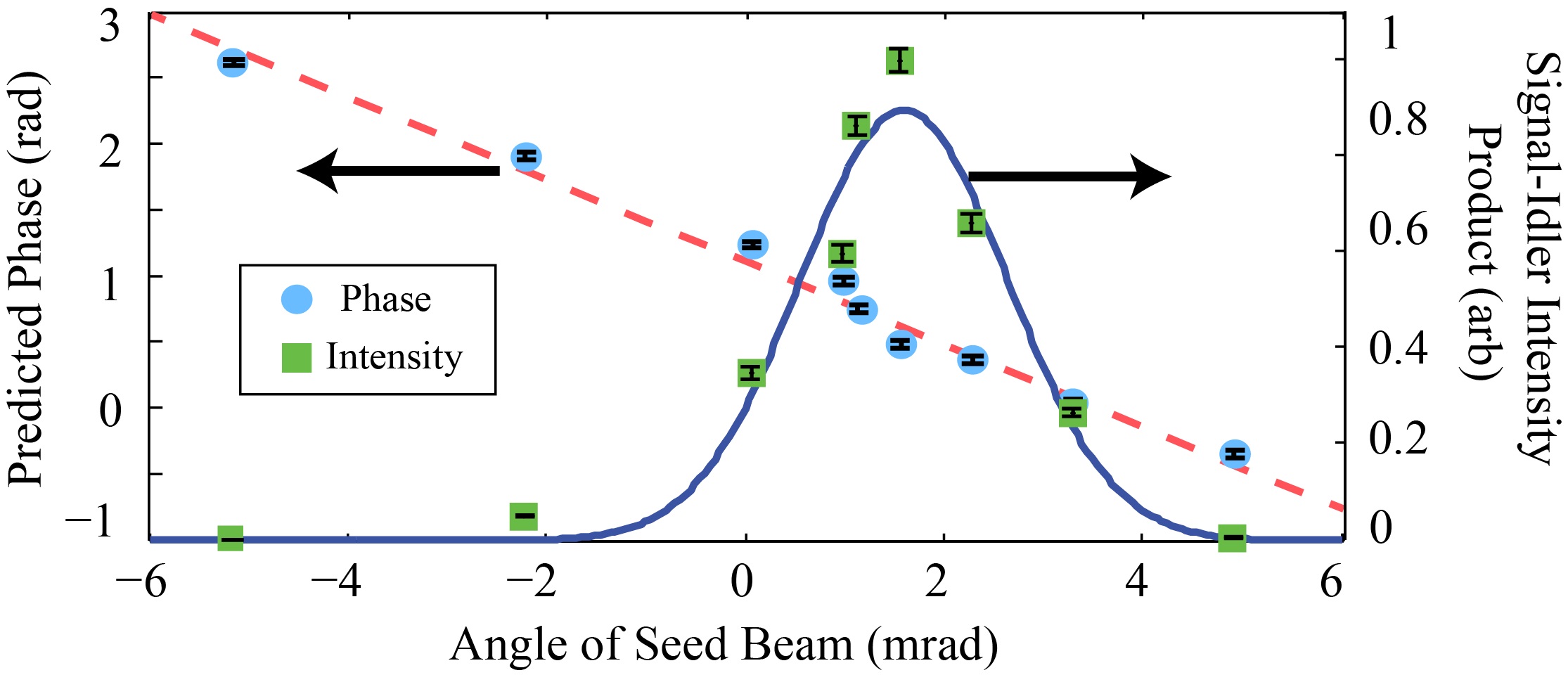}}
\caption{{Effect of the seed angle.} The circles are the phase
of the entangled state predicted by SET plotted versus the seed incidence angle.
 The dashed line is a linear fit to these data, indicating
that the phase changes by 0.312 rad per mrad. The squares are the
product of the experimentally measured intensity of the signal and idler
light that is coupled into single-mode fiber. The solid curve is a
Gaussian fit to these data. }
\end{figure}

In contrast, in SET our seed pulse has a waist of $\omega _{s}\approx 1000$ $%
\mu$m, with a range in transverse momenta of only $\Delta k_{s}\approx 0.3$ mrad.
Thus, one can selectively explore the density matrix of
pairs generated at specific angles, so that SET can easily capture the effect of the phase dependence on the emission angle. Hence SET\ will allow us to
investigate the biphoton wave function that would be generated by SPDC in
even more detail than the usual, emission-angle averaged QST. In fact, the disagreement with QST (as in
comparing Fig. 3a and Fig. 3b) can be understood as the ability of SET to
look ``deeper'' into the biphoton wave function than standard QST,
and actually study the entanglement between polarization and
emission angle. \ It should be possible to obtain the nominal QST matrix
shown in Fig. 3a by using SET, scanning over the same angular ranges, and
averaging the results. \ We will return to these kinds of characterizations
in a future publication; in fact, it has already been reported \cite%
{eckstein_direct_2013,Fang_fast_2014} that the extraction of frequency correlations of
entangled photons that would be observed in a spontaneous experiment can be
done with a much higher resolution in a stimulated experiment.

In conclusion, we have experimentally demonstrated that sources of entangled
photons generated by SPDC can be characterized using a technique based on
stimulated emission, "stimulated emission tomography" (SET). This allowed us
to perform a sort of virtual tomography of the quantum correlated pairs that
would be generated were the stimulating seed beam  absent. 
Especially in low-count-rate sources, stimulated emission tomography should
allow for a faster, less demanding, and more accurate characterization of
sources of entangled photons-pairs than standard quantum state tomography (QST). A
high fidelity between the quantum state deduced from SET and that deduced
from QST was found. \ Differences between
the results of SET and QST can be understood as arising from the emission-angle
averaging that results in usual QST.
Using SET it should be possible to reveal
the underlying structure of the polarization-angle correlations, revealing
an entanglement between polarization and emission-angle to which a usual
application of QST is blind. 

We thank Alan Stummer for his work on our coincidence circuit.
We acknowledge finacial support from the Natural Sciences and Engineering Research Council of Canada (NSERC), and the Canadian Institute for Advances Studies (CIFAR).  

\bibliography{stimulated_tomography}

\begin{thebibliography}{10}
\expandafter\ifx\csname url\endcsname\relax
  \def\url#1{\texttt{#1}}\fi
\expandafter\ifx\csname urlprefix\endcsname\relax\def\urlprefix{URL }\fi
\providecommand{\bibinfo}[2]{#2}
\providecommand{\eprint}[2][]{\url{#2}}

\bibitem{nielsen_quantum_2000}
\bibinfo{author}{Nielsen, M.~A.} \& \bibinfo{author}{Chuang, I.~L.}
\newblock \emph{\bibinfo{title}{Quantum Computation and Quantum Information}}
  (\bibinfo{publisher}{Cambridge University Press}, \bibinfo{year}{2000}),
  \bibinfo{edition}{1} edn.

\bibitem{Raussendorf_measurement_2003}
\bibinfo{author}{Raussendorf, R.}, \bibinfo{author}{Browne, D.~E.} \&
  \bibinfo{author}{Briegel, H.~J.}
\newblock \bibinfo{title}{Measurement-based quantum computation on cluster
  states}.
\newblock \emph{\bibinfo{journal}{Phys. Rev. A}} \textbf{\bibinfo{volume}{68}},
  \bibinfo{pages}{022312} (\bibinfo{year}{2003}).

\bibitem{walther_experimental_2005}
\bibinfo{author}{Walther, P.} \emph{et~al.}
\newblock \bibinfo{title}{Experimental one-way quantum computing}.
\newblock \emph{\bibinfo{journal}{Nature}} \textbf{\bibinfo{volume}{434}},
  \bibinfo{pages}{169--176} (\bibinfo{year}{2005}).

\bibitem{boto_quantum_2000}
\bibinfo{author}{Boto, A.~N.} \emph{et~al.}
\newblock \bibinfo{title}{Quantum interferometric optical lithography:
  Exploiting entanglement to beat the diffraction limit}.
\newblock \emph{\bibinfo{journal}{Physical Review Letters}}
  \textbf{\bibinfo{volume}{85}}, \bibinfo{pages}{2733--2736}
  (\bibinfo{year}{2000}).

\bibitem{mitchell_super-resolving_2004}
\bibinfo{author}{Mitchell, M.~W.}, \bibinfo{author}{Lundeen, J.~S.} \&
  \bibinfo{author}{Steinberg, A.~M.}
\newblock \bibinfo{title}{Super-resolving phase measurements with a multiphoton
  entangled state}.
\newblock \emph{\bibinfo{journal}{Nature}} \textbf{\bibinfo{volume}{429}},
  \bibinfo{pages}{161--164} (\bibinfo{year}{2004}).

\bibitem{Ekert_quantum_1991}
\bibinfo{author}{Ekert, A.~K.}
\newblock \bibinfo{title}{Quantum cryptography based on bell’s theorem}.
\newblock \emph{\bibinfo{journal}{Phys. Rev. Lett.}}
  \textbf{\bibinfo{volume}{67}}, \bibinfo{pages}{661--663}
  (\bibinfo{year}{1991}).

\bibitem{ursin_entanglement_2007}
\bibinfo{author}{Ursin, R.} \emph{et~al.}
\newblock \bibinfo{title}{Entanglement-based quantum communication over 144
  km}.
\newblock \emph{\bibinfo{journal}{Nature physics}}
  \textbf{\bibinfo{volume}{3}}, \bibinfo{pages}{481--486}
  (\bibinfo{year}{2007}).

\bibitem{Jennewein_quantum_2000}
\bibinfo{author}{Jennewein, T.} \emph{et~al.}
\newblock \bibinfo{title}{Quantum cryptography with entangled photons}.
\newblock \emph{\bibinfo{journal}{Phys. Rev. Lett.}}
  \textbf{\bibinfo{volume}{84}}, \bibinfo{pages}{4729--4732}
  (\bibinfo{year}{2000}).

\bibitem{james_measurement_2001}
\bibinfo{author}{James, D. F.~V.} \emph{et~al.}
\newblock \bibinfo{title}{Measurement of qubits}.
\newblock \emph{\bibinfo{journal}{Phys. Rev. A}} \textbf{\bibinfo{volume}{64}},
  \bibinfo{pages}{052312} (\bibinfo{year}{2001}).

\bibitem{gill_state_2000}
\bibinfo{author}{Gill, R.~D.} \& \bibinfo{author}{Massar, S.}
\newblock \bibinfo{title}{State estimation for large ensembles}.
\newblock \emph{\bibinfo{journal}{Phys. Rev. A}} \textbf{\bibinfo{volume}{61}},
  \bibinfo{pages}{042312} (\bibinfo{year}{2000}).

\bibitem{Okamoto_experimental_2012}
\bibinfo{author}{Okamoto, R.} \emph{et~al.}
\newblock \bibinfo{title}{Experimental demonstration of adaptive quantum state
  estimation}.
\newblock \emph{\bibinfo{journal}{Phys. Rev. Lett.}}
  \textbf{\bibinfo{volume}{109}}, \bibinfo{pages}{130404}
  (\bibinfo{year}{2012}).

\bibitem{mahler_adaptive_2013}
\bibinfo{author}{Mahler, D.~H.} \emph{et~al.}
\newblock \bibinfo{title}{Adaptive quantum state tomography improves accuracy
  quadratically}.
\newblock \emph{\bibinfo{journal}{Phys. Rev. Lett.}}
  \textbf{\bibinfo{volume}{111}}, \bibinfo{pages}{183601}
  (\bibinfo{year}{2013}).

\bibitem{gross_quantum_2010}
\bibinfo{author}{Gross, D.} \emph{et~al.}
\newblock \bibinfo{title}{Quantum state tomography via compressed sensing}.
\newblock \emph{\bibinfo{journal}{Phys. Rev. Lett.}}
  \textbf{\bibinfo{volume}{105}}, \bibinfo{pages}{150401}
  (\bibinfo{year}{2010}).

\bibitem{Liscidini_Stimulated_2013}
\bibinfo{author}{Liscidini, M.} \& \bibinfo{author}{Sipe, J.~E.}
\newblock \bibinfo{title}{Stimulated emission tomography}.
\newblock \emph{\bibinfo{journal}{Phys. Rev. Lett.}}
  \textbf{\bibinfo{volume}{111}}, \bibinfo{pages}{193602}
  (\bibinfo{year}{2013}).

\bibitem{horn_Monolithic_2012}
\bibinfo{author}{Horn, R.} \emph{et~al.}
\newblock \bibinfo{title}{Monolithic source of photon pairs}.
\newblock \emph{\bibinfo{journal}{Phys. Rev. Lett.}}
  \textbf{\bibinfo{volume}{108}}, \bibinfo{pages}{153605}
  (\bibinfo{year}{2012}).

\bibitem{orieux_direct_2013}
\bibinfo{author}{Orieux, A.} \emph{et~al.}
\newblock \bibinfo{title}{Direct bell states generation on a iii-v
  semiconductor chip at room temperature}.
\newblock \emph{\bibinfo{journal}{Phys. Rev. Lett.}}
  \textbf{\bibinfo{volume}{110}}, \bibinfo{pages}{160502}
  (\bibinfo{year}{2013}).

\bibitem{horn_inherent_2013}
\bibinfo{author}{Horn, R.~T.} \emph{et~al.}
\newblock \bibinfo{title}{Inherent polarization entanglement generated from a
  monolithic semiconductor chip}.
\newblock \emph{\bibinfo{journal}{Scientific reports}}
  \textbf{\bibinfo{volume}{3}} (\bibinfo{year}{2013}).

\bibitem{eckstein_direct_2013}
\bibinfo{author}{Eckstein, A.} \emph{et~al.}
\newblock \bibinfo{title}{High-resolution spectral characterization of two
  photon states via classical measurements}.
\newblock \emph{\bibinfo{journal}{Laser and Photonics Reviews}}
  \textbf{\bibinfo{volume}{8}}, \bibinfo{pages}{L76--L80}
  (\bibinfo{year}{2014}).

\bibitem{Fang_fast_2014}
\bibinfo{author}{Fang, B.} \emph{et~al.}
\newblock \bibinfo{title}{Fast and highly resolved capture of the joint
  spectral density of photon pairs}.
\newblock \emph{\bibinfo{journal}{Optica}} \textbf{\bibinfo{volume}{1}},
  \bibinfo{pages}{281--284} (\bibinfo{year}{2014}).

\bibitem{Kwiat_typeII_1995}
\bibinfo{author}{Kwiat, P.~G.} \emph{et~al.}
\newblock \bibinfo{title}{New high-intensity source of polarization-entangled
  photon pairs}.
\newblock \emph{\bibinfo{journal}{Phys. Rev. Lett.}}
  \textbf{\bibinfo{volume}{75}}, \bibinfo{pages}{4337--4341}
  (\bibinfo{year}{1995}).

\bibitem{altepeter_Phase-compensated_2005}
\bibinfo{author}{Altepeter, J.}, \bibinfo{author}{Jeffrey, E.} \&
  \bibinfo{author}{Kwiat, P.}
\newblock \bibinfo{title}{Phase-compensated ultra-bright source of entangled
  photons}.
\newblock \emph{\bibinfo{journal}{Opt. Express}} \textbf{\bibinfo{volume}{13}},
  \bibinfo{pages}{8951--8959} (\bibinfo{year}{2005}).

\bibitem{Rangarajan_optimizing_2009}
\bibinfo{author}{Rangarajan, R.}, \bibinfo{author}{Goggin, M.} \&
  \bibinfo{author}{Kwiat, P.}
\newblock \bibinfo{title}{Optimizing type-i polarization-entangled photons}.
\newblock \emph{\bibinfo{journal}{Opt. Express}} \textbf{\bibinfo{volume}{17}},
  \bibinfo{pages}{18920--18933} (\bibinfo{year}{2009}).

\bibitem{ling_absolute_2008}
\bibinfo{author}{Ling, A.}, \bibinfo{author}{Lamas-Linares, A.} \&
  \bibinfo{author}{Kurtsiefer, C.}
\newblock \bibinfo{title}{Absolute emission rates of spontaneous parametric
  down-conversion into single transverse gaussian modes}.
\newblock \emph{\bibinfo{journal}{Phys. Rev. A}} \textbf{\bibinfo{volume}{77}},
  \bibinfo{pages}{043834} (\bibinfo{year}{2008}).

\bibitem{jozsa_fidelity_1994}
\bibinfo{author}{Jozsa, R.}
\newblock \bibinfo{title}{Fidelity for mixed quantum states}.
\newblock \emph{\bibinfo{journal}{Journal of Modern Optics}}
  \textbf{\bibinfo{volume}{41}}, \bibinfo{pages}{2315--2323}
  (\bibinfo{year}{1994}).

\bibitem{SNLO}
\bibinfo{note}{{SNLO} nonlinear optics code available from {A. V. Smith},
  {AS-Photonics, Albuquerque, NM}}.

\end{thebibliography}
\cleardoublepage

\section*{Supplement}

\section{Data Normalization and Fitting}
In this section we describe how the intensities that we measured in the lab were used to reconstruct a least-squares estimate of the two-photon density matrix.
In standard two-photon polarization quantum state tomography (QST), at least 16 (but typically 36) different separable projective measurements are performed on the two photons.
From these measurements, the probability of projecting the two-photon state $\rho$ onto each projector is estimated\cite{james_measurement_2001}.
In a typical experiment (Fig. 2b), each photon is sent to a different polarizing beamsplitter (PBS) and then the rates of both photons being transmitted ($R_{HH}$), both photons being reflected ($R_{VV}$), the first being transmitted and the second reflected ($R_{HV}$), and the second being transmitted and the first reflected ($R_{VH}$) are measured and used to estimate the probabilities of the different outcomes $P_{HH}$, $P_{HV}$, $P_{VH}$, and $P_{VV}$ as $P_{HH}=R_{HH}/(R_{HH}+R_{HV}+R_{VH}+R_{VV})$ etc.
These probabilities are taken to correspond to projecting $\rho$ onto the projection operators $\ket{H}\bra{H}\otimes\ket{H}\bra{H}$ etc.
This is then repeated for several other measurement settings, and a least-squares estimate of $\rho$, $\tilde{\rho}_\mathrm{LS}$ is constructed by minimizing a cost function
\begin{equation}
\label{eq:ls}
C=\sum_{s,i} \left( \mathrm{Tr}\left[ \hat{O}_{s,i} \hat{\rho}_\mathrm{LS} \right] - P_{s,i} \right)^2
\end{equation}
over $\hat{\rho}_\mathrm{LS}$.
In equation \ref{eq:ls}, $s$ and $i$ label the signal and idler polarizations; i.e.
\begin{equation}
P_{s,i}= \left\{P_{HH}, P_{HV}, P_{VH}, P_{VV}, P_{DD}, \dots \right\},
\end{equation}
and
\begin{equation}
\hat{O}_{s,i}=\left\{\ket{HH}\bra{HH},\ket{HV}\bra{HV}, \dots \right\}
\end{equation}
(using $\ket{HH}\bra{HH}$ as shorthand for $\ket{H}\bra{H}\otimes\ket{H}\bra{H}$).

In stimulated emission tomography (SET), these probabilities $P_{s,i}$ are estimated by comparing the stimulated intensity to the seed intensity.
Ideally, this is done by seeding with different polarizations and measuring the intensity of the the stimulated light in different polarization bases.
Then these stimulated intensities are normalized by the seed intensity to estimate probabilities.
For example, if the seed diagonally polarized, and the stimulated beam is measured in the horizontal basis we can compute $P_{D,H}$ (the probability of detecting a diagonally-polarized photon and a horizontally-polarized photon in a spontaneous experiment) as:
\begin{equation}\label{eq:idealProb}
P_{D,H}^\mathrm{ideal} = \frac{I^\mathrm{stim}_H}{I^\mathrm{seed}_D},
\end{equation}
where $I^\mathrm{stim}_H$ is the intensity of the simulated light detected at the transmitted port of $\mathrm{PBS}_i$ in Fig. 2c, and $I^\mathrm{seed}_D$ is the total intensity of the seed light that is coupled into the signal fibre.
(Note we measure $I^\mathrm{seed}_D$ by summing the intensity that is measured exiting both output ports of $\mathrm{PBS}_s$ in Fig. 2c for any waveplate setting).

In practice, however, the coupling efficiencies of the signal and idler modes can be different so $\frac{I^\mathrm{stim}_H}{I^\mathrm{seed}_D}$ will not always result in the ``true probability'' $P_{D,H}$.
For example, if the idler mode is coupled with efficiency $\epsilon_i$, and the signal mode with efficiency $\epsilon_s$ then  
\begin{equation}\label{eq:idealProb2}
P_{D,H}^\mathrm{ideal} = \frac{I^\mathrm{stim}_H}{I^\mathrm{seed}_D}=\frac{\epsilon_i}{\epsilon_s}\times P_{DH},
\end{equation}
which is proportional to the true probability, but has an additional ``coupling factor''.
To remedy this we take an additional step in our normalization.
We first compute the {ideal probabilities} as defined by equation \ref{eq:idealProb2} for all seed polarizations and stimulated measurement combinations.
Then we renormalize these probabilities as 
\begin{equation} 
\tilde{P}_{DH}=\frac{{P}_{DH}^\mathrm{ideal}}{{P}_{DH}^\mathrm{ideal}+{P}_{DV}^\mathrm{ideal}+{P}_{AH}^\mathrm{ideal}+{P}_{AV}^\mathrm{ideal}}.
\end{equation}
Now, since the {ideal probabilities} are all proportional to $\frac{\epsilon_i}{\epsilon_s}$, this coupling factor will cancel out and we are left with 
\begin{equation} 
\tilde{P}_{DH}=\frac{{P}_{DH}}{{P}_{DH}+{P}_{DV}+{P}_{AH}+{P}_{AV}},
\end{equation}
and since ${P}_{DH}+{P}_{DV}+{P}_{AH}+{P}_{AV}=1$, we have
\begin{equation}
\tilde{P}_{DH}={{P}_{DH}}.
\end{equation}
Thus even with different signal and idler coupling efficiencies we can extract the true probabilities.

In our experiment, we find one other systematic error.
Namely, if we prepare diagonally-polarized seed light, the light that is ultimately coupled into the signal mode is not diagonally polarized.
For example, we observe a birefringent phase shift, so that if we prepare the seed in ($\ket{H}+\ket{V})/\sqrt{2}$ we find that the light coupled in the signal fiber has been rotated to $(\ket{H}+e^{i\phi}\ket{V})/\sqrt{2}$).
In addition there is a polarization-dependent coupling efficiency, by which we mean that $\ket{H}+\ket{V})/\sqrt{2}$ will go to $|a|\ket{H}+|b|\ket{V}$ with $|a|\neq |b|$).
To characterize and correct for this, we reconstruct the polarization state of the seed light coupled that is coupled into the signal fiber, using single-photon QST.
Then we perform our least-squared fit with a rotated basis.
In other words, when we seed with (say) diagonally-polarized light and then find that this results in a different polarization $\rho_{D'}$ coupled into the signal fiber, we change the corresponding measurement operator from $\hat{O}_{D,i}=\ket{D}\bra{D}\otimes\ket{i}\bra{i}$ to  $\hat{O}_{D',i}=\rho_{D'}\otimes\ket{i}\bra{i}$.
We do this for all six seed polarizations (H, V, D, A, R, and L), measuring the stimulated light in the same six bases for each seed polarization.
A representative example of these seed polarization reconstructions is shown in Table 1.
Then the 36 different probabilities and rotated measurement operators are fed into our least-squares fitting routine to minimize equation \ref{eq:ls}.
As described in the main text, we find that this procedure predicts the amplitudes, purity and concurrence of the entangled state very well.

\section{Signal-Idler Polarization Correlations}

One way to understand stimulated-emission tomography, is in terms of polarization correlations between the seed signal beam and the stimulated idler beam.
For example, if an entangled photon-pair source is configured to produce the state $(\ket{HH}_{s,i}+\ket{VV}_{s,i})/\sqrt{2}$ and it is stimulated, then the polarization correlations listed in Table 1 will be observed.
Namely, if the signal beam is diagonally polarized the idler beam will also be diagonally polarized, while if the signal beam is right-circularly polarized the idler beam will be left-circularly polarized, etc.
However if the source is misaligned, so that it creates an equal mixture of $\ket{HH}$ and $\ket{VV}$ (rather than a quantum superposition) then if the signal beam is diagonally or right-circularly polarized the idler beam will always be incoherently polarized (i.e. a mixture of $\ket{H}$ and $\ket{V}$).
Other variations of the spontaneously-generated two-photon state map on to correlations between the signal and idler polarizations as expected.
A non-zero phase between $\ket{HH}$ and $\ket{VV}$ results in a rotation of the stimulated idler polarizations, and unequal amplitudes of $\ket{HH}$ and $\ket{VV}$ map onto unequal amplitudes of $\ket{H}$ and $\ket{V}$ in the idler polarization.

\section{Polarization-Dependent Loss}
In this section, we address an important experimental consideration:
the theory of \cite{Liscidini_Stimulated_2013}, and all of the discussion above implicitly assumes a single spatial mode, i.e. all of the seed light incident in the signal mode stimulates light into the idler mode, and the light in both modes is detected.
In the data presented above this was approximately true; however, it is possible for this assumption to break down.
In our experiment, this breakdown occurs if the ``seed lens'' (Fig. 2) is removed so that the incoming seed beam is approximately the same size as the pump beam.
In this configuration we find that it is possible for the seed beam to stimulate light into the idler mode, and then to be ``walked out'' of the signal mode so that it is not detected.
The opposite could also happen -- light could be coupled into the signal mode without stimulating any light into the idler mode.
Such errors lead to discrepancies between the predictions of SET and QST.

The symptoms and result of this problem are as follows. 
When a diagonally polarized seed beam is incident, we would expect that the ratio of the intensity of the horizontally-polarized light to the vertically-polarized light that is coupled into the signal fibre is $R_{s_H/s_V}\approx 1$.
However, due to birefringent spatial walk-off, either horizontally or vertically polarized light could be coupled into the signal mode more efficiently.
This would depend on the incidence angle of the seed beam ($\theta$ in Fig. 2 of the main text).
In our experiment when the seed beam waist is approximately the same size as the pump beam waist, we find that $R_{s_H/s_V}$ can vary from $0.1$ to $10$ when $\theta$ varies by approximately $10$ mrad.
In any case, if the spontaneously-generated two-photon state is close to $(\ket{HH}+\ket{VV})/\sqrt{2}$,
then ideally the ratio of the horizontal intensity to vertical intensity for the idler polarization $R_{i_H/i_V}$ would follow the seed polarization, i.e. $R_{i_H/i_V} = R_{s_H/s_V}$ (even if $R_{s_H/s_V}\neq 1$).  
But we find without the seed lens this is not the case, and so, when the waist of the seed beam is small, the results of SET can vary a great deal.

For example, in one configuration (when the source was configured to produce $(\ket{HH}+\ket{VV})/\sqrt{2}$) we found $R_{s_H/s_V}=10.5$ and $R_{i_H/i_V}=11.6$.
In this case the predictions of SET and QST agree: SET predicts a concurrence of $0.915$, while QST predicts a concurrence of $0.916$.
However, changing $\theta$ by a few mrad resulted in $R_{s_H/s_V}=1.16$ and $R_{i_H/i_V}=0.064$.
In this setting SET predicts a concurrence of $0.442$ -- greatly underestimating the ``true concurrence'' of $0.916$ revealed by QST (which does not change when the seed beam changes).
This illustrates that care must be taken so that the assumptions made in \cite{Liscidini_Stimulated_2013} are met.
In our experiment, this meant making the seed waist larger than the birefringent spatial walk-off.
In other systems this particular source of error may not be as prevalent, but care must still be taken to ensure that these sorts of errors are not occurring in other modes, such as temporal or frequency modes.
Note that making a single-photon coincidence measurement to estimate the actual ratio of the amplitudes of $\ket{HH}$ to $\ket{VV}$ and comparing to this result to $R_{s_H/s_V}$ and $R_{i_H/i_V}$ would immediately reveal this error.
We checked that this was the case for all of our data presented in the paper.
However, we stress that in our experiment the seed waist was large, so we were not sensitive to this error.
With a large seed waist, $\theta$ must to be changed by an amount that would appreciably decrease the intensity of either the signal or idler beam before $R_{i_H/i_V} \neq R_{s_H/s_V}$.

\begin{table*}[htbp]
\centering
\caption{A list of the polarization states that are stimulated into the idler mode (second column) for each seed polarization state (first column) when the SPDC source is configured to produce the nominal state $(\ket{HH}_{s,i}+\ket{VV}_{s,i})/\sqrt{2}$.  Example experimentally reconstructed seed and stimulated polarization states are also shown.}
\begin{tabular}{cc|cc}
\hline
\multicolumn{2}{c|}{Seeding Signal Polarizations}&
\multicolumn{2}{c}{Stimulated Idler Polarizations}\\
Ideal & Experimental & Ideal & Experimental\\
\hline
$\ket{H} = 
\left(\begin{array}{cc}
1      & 0 \\
0      & 0 
\end{array}\right)$ & 
$\left(\begin{array}{cc}
0.996&-0.020+0.058i \\
-0.020-0.058i&0.004
\end{array}\right)$ & 
$\ket{H}$ &
$\left(\begin{array}{cc}
0.936   &  -0.062-0.053i\\
  -0.062+0.053i  &   0.064 
\end{array}\right)$\\
$\ket{V} = 
\left(\begin{array}{cc}
0      & 0 \\
0      & 1 
\end{array}\right)$ & 
$\left(\begin{array}{cc}
0.002   &  0.025-0.031i\\
   0.025+0.031i  &  0.998
\end{array}\right)$ & 
$\ket{V}$ &
$\left(\begin{array}{cc}
0.016   & -0.026+0.024i\\
  -0.026-0.024i   &   0.984 
\end{array}\right)$\\
$\ket{D} = 
\left(\begin{array}{cc}
0.5      & 0.5 \\
0.5     & 0.5 
\end{array}\right)$ & 
$\left(\begin{array}{cc}
   0.506     &  0.492-0.068i\\
   0.492+0.068i    &  0.494
\end{array}\right)$ & 
$\ket{D}$ &
$\left(\begin{array}{cc}
0.514  &    0.482-0.056i\\
   0.482+0.056i  &   0.486
\end{array}\right)$\\
$\ket{A} = 
\left(\begin{array}{cc}
0.5      & -0.5 \\
-0.5     & 0.5 
\end{array}\right)$ & 
$\left(\begin{array}{cc}
   0.449   & -0.484+0.107i\\
  -0.484-0.107i   &   0.551
\end{array}\right)$ & 
$\ket{A}$ &
$\left(\begin{array}{cc}
0.438 &   -0.485+0.076i  \\
  -0.485-0.076i  &    0.562
\end{array}\right)$\\
$\ket{R} = 
\left(\begin{array}{cc}
0.5      & -0.5i \\
0.5     & 0.5 
\end{array}\right)$ & 
$\left(\begin{array}{cc}
0.525   &  -0.080-0.489i\\
  -0.080+0.489i   &  0.475
\end{array}\right)$ & 
$\ket{L}$ &
$\left(\begin{array}{cc}
0.540    &  0.011+0.498i  \\
   0.011-0.498i   &    0.460
\end{array}\right)$\\
$\ket{L} = 
\left(\begin{array}{cc}
0.5      & 0.5i \\
-0.5     & 0.5 
\end{array}\right)$ & 
$\left(\begin{array}{cc}
0.430   &   0.081+0.484i\\
   0.081-0.484i   &  0.570
\end{array}\right)$ & 
$\ket{R}$ &
$\left(\begin{array}{cc}
0.419 &   -0.080-0.433i\\
  -0.080+0.433i   &  0.581
\end{array}\right)$\\
\hline
\end{tabular}
\end{table*}

\end{document}